\documentclass[conference]{IEEEtran}

\usepackage[colorlinks]{hyperref}
\usepackage{makeidx}
\makeindex

\usepackage{amsmath}
\usepackage{amsthm}
\usepackage{amssymb}
\usepackage{bm}
\usepackage{xspace}
\usepackage{xcolor}
\usepackage{graphicx}
\usepackage{url}
\usepackage{framed}
\usepackage{float}
\usepackage{rotating}
\usepackage{verbatim}
\usepackage{listings}
\usepackage{lscape}

\usepackage{pgfplots}
\usepackage{pgf}
\usepackage{tikz}
\usetikzlibrary{arrows,shapes.misc,chains,scopes}
\pgfplotsset{compat = newest}
\usepackage{pgfplotstable}
\usepackage{booktabs}
\usepackage{colortbl}

\newcommand{\executeiffilenewer}[3]{%
\ifnum\pdfstrcmp{\pdffilemoddate{#1}}%
{\pdffilemoddate{#2}}>0%
{\immediate\write18{#3}}\fi%
}

\graphicspath{{figures/}}

\theoremstyle{plain}
\newtheorem{proposition}{Proposition}

\newcounter{algocount}
\newcounter{examplecount}

\newenvironment{algorithm}[1][]{\refstepcounter{algocount}\begin{trivlist}\item \textbf{Algorithm \thealgocount.}#1\\[-0.2cm]\rule{\columnwidth}{1pt}}{\\[-0.2cm]\rule{\columnwidth}{1pt}\end{trivlist}}

\newcommand{\setr}{\ensuremath{\mathbf{R}}\xspace}

\DeclareMathOperator*{\argmin}{argmin}
\DeclareMathOperator*{\argmax}{argmax}

\DeclareMathOperator{\expop}{\mathbb{E}}
\DeclareMathOperator{\entop}{\mathbb{H}}

\DeclareMathOperator{\kl}{\mathbb{D}}

\usepackage{cite}
\usepackage{printlen}
\usepackage{booktabs}

\newcommand{\bsm}[1]{\begin{array}{#1}}
\newcommand{\esm}{\end{array}}

\title{Fixed-to-Variable Length Distribution Matching}

\IEEEoverridecommandlockouts

\author{\IEEEauthorblockN{Rana Ali Amjad and Georg B\"ocherer}
\IEEEauthorblockA{Institute for Communications Engineering\\Technische Universit\"at M\"unchen, Germany\\
Email: \texttt{raa2463@gmail.com,georg.boecherer@tum.de}}
\thanks{This work was supported by the German Ministry of Education and Research in the framework of an Alexander von Humboldt Professorship.}
}

\DeclareMathOperator{\supp}{supp}

\begin{document}

\maketitle

\begin{abstract} Fixed-to-variable length (f2v) matchers are used to reversibly transform an input sequence of independent and uniformly distributed bits into an output sequence of bits that are (approximately) independent and distributed according to a target distribution. The degree of approximation is measured by the informational divergence between the output distribution and the target distribution. An algorithm is developed that efficiently finds optimal f2v codes. It is shown that by encoding the input bits blockwise, the informational divergence per bit approaches zero as the block length approaches infinity. A relation to data compression by Tunstall coding is established.
\end{abstract}

\section{Introduction}

Distribution matching considers the problem of mapping uniformly distributed bits to symbols that are approximately distributed according to a target distribution. In difference to the simulation of random processes \cite{steinberg1996simulation} or the exact generation of distributions \cite{knuth1976complexity}, distribution matching requires that the original bit sequence can be \emph{recovered} from the generated symbol sequence. We measure the degree of approximation by the normalized informational divergence (I-divergence), which is an appropriate measure when we want to achieve channel capacity of noisy and noiseless channels \cite[Sec.~3.4.3~\&~Chap.~6]{bocherer2012capacity} by using a matcher. A related work is \cite{bocherer2011matching},\cite[Chap. 3]{bocherer2012capacity}, where it is shown that \emph{variable-to-fixed length} (v2f) matching is optimally done by geometric Huffman coding and the relation to \emph{fixed-to-variable length} (f2v) source encoders is discussed. In the present work, we consider binary distribution matching by prefix-free f2v codes.

\subsection{Rooted Trees With Probabilities}
\begin{figure}
\footnotesize
\centering
\def\svgwidth{\columnwidth}
\executeiffilenewer{figures/trees.svg}{figures/trees.pdf}%
{inkscape -z -D --file=figures/trees.svg %
--export-pdf=figures/trees.pdf --export-latex}%
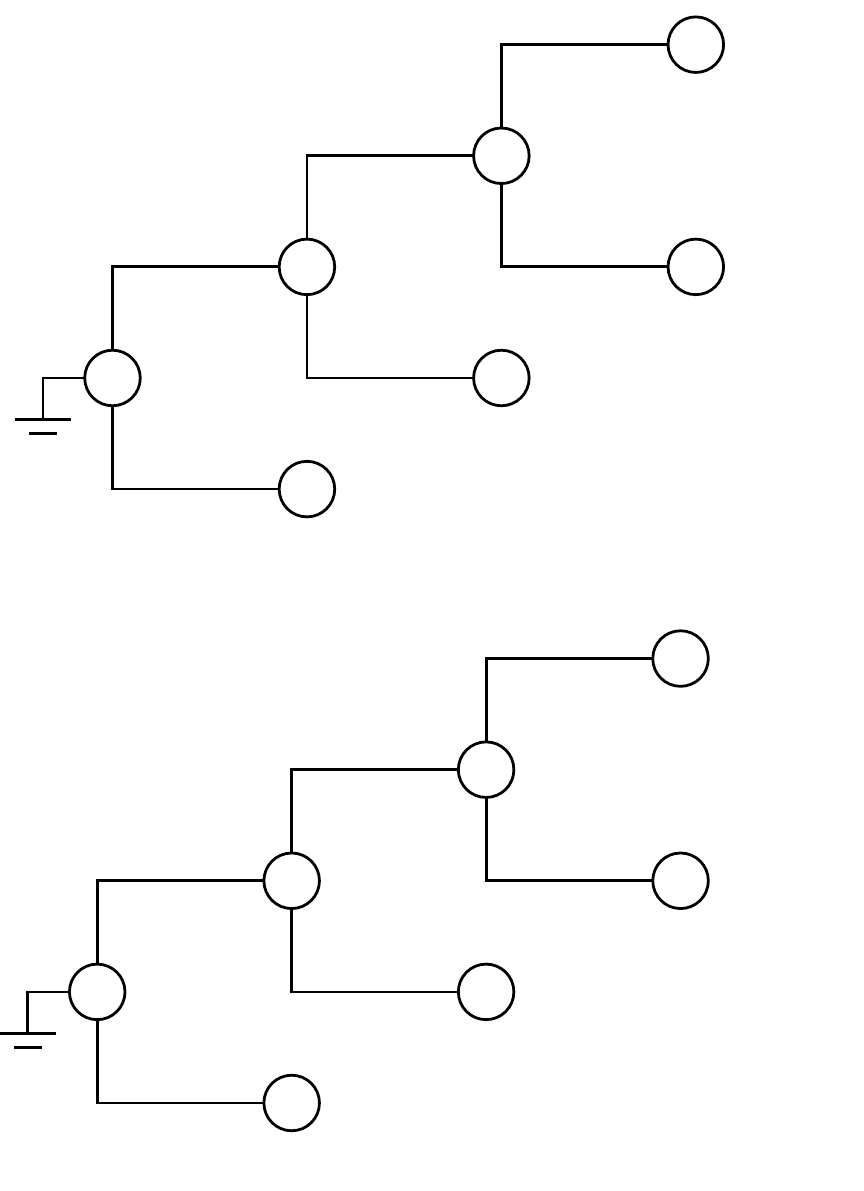%

\vspace{-0.5cm}
\caption{A rooted tree with probabilities. The set of branching nodes is $\mathcal{B}=\{1,2,4\}$ and the set of leaf nodes is $\mathcal{L}=\{3,5,6,7\}$. In (a), a uniform leaf distribution $U_T$ is chosen, i.e., for each $i\in\mathcal{L}$, $U_T(i)=\frac{1}{4}$. The leaf distribution determines the node probabilities and branching distributions. In (b), identical branching distributions are chosen, i.e., for each $i\in\mathcal{B}$, $Q_T^i=Q$, where $Q(0)=:q_0=\frac{2}{3}$ and $Q(1)=:q_1=\frac{1}{3}$. The branching distributions determine the resulting node probabilities, which we denote by $Q_T(i)$. Since the tree is complete, $\sum_{i\in\mathcal{L}}Q_T(i)=1$ and $Q_T$ defines a leaf distribution.
}\label{fig:trees}
\vspace{-0.5cm}
\end{figure}
We use the framework of rooted trees with probabilities \cite{rueppel1994leaf},\cite{bocherer2013rooted}. Let $\mathcal{T}$ be the set of all binary trees with $2^m$ leaves and consider some tree $T\in\mathcal{T}$. Index all nodes by the numbers $\mathcal{N}:=\{1,2,3,\dotsc\}$ where $1$ is the root. Note that there are at least $|\mathcal{N}|\geq 2^{m+1}-1$ nodes in the tree, with equality if the tree is complete. A tree is complete if any right-infinite binary sequence starts with a path from the root to a leaf. Let $\mathcal{L}\subset \mathcal{N}$ be the set of leaf nodes and let  $\mathcal{B}:=\mathcal{N}\setminus\mathcal{L}$ be the set of branching nodes. Probabilities can be assigned to the tree by defining a distribution over the $2^m$ paths through the tree. For each $i\in\mathcal{N}$, denote by $P_T(i)$ the probability that a path is chosen that passes through node $i$. Since each path ends at a different leaf node, $P_T$ defines a leaf distribution, i.e., $\sum_{i\in\mathcal{L}}P_T(i)=1$. For each branching node $i\in\mathcal{B}$, denote by $P_T^i$ the branching distribution, i.e., the probabilities of branch $0$ and branch $1$ after passing through node $i$. The probabilities on the tree are completely defined either by defining the branching distributions $\{P_T^i,i\in\mathcal{B}\}$ or by defining the leaf distribution $\{P_T(i),i\in\mathcal{L}\}$. See Fig.~\ref{fig:trees} for an example.

\subsection{v2f Source Encoding and f2v Distribution Matching}

Consider a binary distribution $Q$ with $q_1=Q(0)$, $q_1=Q(1)$, $0<q_0<1$, and a binary tree $T$ with $2^m$ leaves. Let $Q_T(i)$, $i\in\mathcal{N}$, be the node probabilities that result from having all branching distributions equal to $Q$, i.e. $Q_T^i=Q$ for each $i\in\mathcal{B}$. See Fig.~\ref{fig:trees}(b) for an example. Let $U_T$ be a uniform leaf distribution, i.e., $U_T(i)=2^{-m}$ for each $i\in\mathcal{L}$, see Fig.~\ref{fig:trees}(a) for an example. We use the tree as a v2f source code for a \emph{discrete memoryless source} (DMS) $Q$. To guarantee lossless compression, the tree for a v2f source encoder has to be complete. Consequently, $Q_T$ defines a leaf distribution, i.e., $\sum_{i\in\mathcal{L}}Q_T(i)=1$. We denote the set of complete binary trees with $2^m$ leaves by $\mathcal{C}$. Each code word consists of $\log_2 2^m=m$ bits and the resulting entropy rate at the encoder output is
\begin{align}
&\frac{1}{m}\entop(Q_T)=\sum_{i\in\mathcal{L}}Q_T(i)[-\log_2 Q_T(i)]\nonumber\\
&=\frac{1}{m}\sum_{i\in\mathcal{L}}Q_T(i)[-\log_2 Q_T(i)+\log_2 U_T(i)-\log_2 U_T(i)]\nonumber	\\
&=1 - \frac{1}{m}\kl(Q_T\Vert U_T)\label{eq:compressionobjective}
\end{align}
where $\entop(Q_T)$ is the entropy of the leaf distribution defined by $Q_T$ and where $\kl(Q_T\Vert U_T)$ is defined accordingly. From \eqref{eq:compressionobjective}, we conclude that the objective is to solve
\begin{align}
\min_{T\in\mathcal{C}}\kl(Q_T\Vert U_T).\label{eq:sourcecodingproblem}
\end{align}
The solution is known to be attained by Tunstall coding \cite{tunstall1967synthesis}. The tree in Fig.~\ref{fig:trees} is a Tunstall code for $Q\colon q_0=\frac{2}{3}$, $q_1=\frac{1}{3}$ and $m=2$ and the corresponding v2f source encoder is
\begin{align}
000\mapsto 00,\;001\mapsto 01,\;01\mapsto 10,\;1\mapsto 11.
\end{align}
The dual problem is f2v distribution matching. $Q$ is now a binary target distribution and we generate the codewords defined by the paths through a (not necessarily complete) binary tree uniformly according to $U_T$. For example, the f2v distribution matcher defined by the tree in Fig.~\ref{fig:trees} is
\begin{align}
00\mapsto 000,\;01\mapsto 001,\;10\mapsto 01,\;11\mapsto 1.
\end{align}
Denote by $\ell_i,i\in\mathcal{L}$ the path lengths and let $L$ be a random variable that is uniformly distributed over the path lengths according to $U_T$. We want the I-divergence per output bit of $U_T$ and $Q_T$ to be small, i.e., we want to solve
\begin{align}
\min_{T\in\mathcal{T}}\frac{\kl(U_T\Vert Q_T)}{\expop_{U_T}(L)}.\label{eq:f2v}
\end{align}
In contrast to \eqref{eq:sourcecodingproblem}, the minimization is now over the set of all (not necessarily complete) binary trees with $2^m$ leaves. Note that although for a non-complete tree we have $\sum_{i\in\mathcal{L}}Q_T(i)<1$, the problem \eqref{eq:f2v} is well-defined, since there is always a complete tree with leaves $\mathcal{L}'\supseteq\mathcal{L}$ and $\sum_{i\in\mathcal{L}'}Q_T(i)=1$. The sum in \eqref{eq:f2v} is over the support of $U_T$, which is $\mathcal{L}$. Solving \eqref{eq:f2v} is the problem that we consider in this work.
\subsection{Outline}
In Sec.~\ref{sec:infodiv} and Sec.~\ref{sec:ninfodiv}, we restrict attention to complete trees. We show that Tunstall coding applied to $Q$ minimizes $\kl(U_T\Vert Q_T)$ and that iteratively applying Tunstall coding to weighted versions of $Q$ minimizes $\kl(U_T\Vert Q_T)/\expop_{U_T}(L)$. In Sec.~\ref{sec:complete} we derive conditions for the optimality of complete trees and show that the I-divergence per bit can be made arbitrarily small by letting the blocklength $m$ approach infinity. Finally, in Sec.~\ref{sec:example}, we illustrate by an example that source decoders are sub-optimal distribution matchers and vice-versa, distribution dematchers are sub-optimal source encoders.
\section{Minimizing I-divergence}\label{sec:infodiv}
Let $\setr$ be the set of real numbers. For a finite set $\mathcal{S}$, we say that $W\colon \mathcal{S}\to\setr$ is a \emph{weighted distribution} if for each $i\in\mathcal{S}$, $W(i)> 0$. We allow for $\sum_{i\in\mathcal{S}}W(i)\neq 1$. The I-divergence of a distribution $P$ and a weighted distribution $W$ is
\begin{align}
\kl(P\Vert W)=\sum_{i\in\supp P}P(i)\log_2\frac{P(i)}{W(i)}
\end{align}
where $\supp$ denotes the support of $P$. The reason why we need this generalization of the notion of distributions and I-divergence will become clear in the next section.
\begin{proposition}\label{prop:infodiv}
Let $Q$ be a weighted binary target distribution, and let
\begin{align}
T^*=\argmin_{T\in\mathcal{C}}\kl(U_T\Vert Q_T)
\end{align}
be an optimal complete tree. Then we find that
\begin{itemize}
\item[i.] An optimal complete tree $T^*$ can be constructed by applying Tunstall coding to $Q$.
\item[ii.] If $0\leq q_0\leq 1$ and $0\leq q_1\leq 1$, then $T^*$ also minimizes $\kl(U_T\Vert Q_T)$ among all possibly non-complete binary trees $\mathcal{T}$, i.e., the optimal tree is complete.
\end{itemize}
\end{proposition}
\begin{IEEEproof}\emph{Part i.} 
We write
\begin{align}
\kl(U_T\Vert Q_T) &= \sum_{i\in\mathcal{L}} 2^{-m}\log_2 \frac{2^{-m}}{Q_T(i)}\nonumber\\
&=-m-2^{-m}\sum_{i \in \mathcal{L}} \log Q_T(i) \label{eq:denominator}
\end{align}
and hence
\begin{equation}
\argmin_{T\in\mathcal{C}} \kl(U_T\Vert Q_T)  =  \argmax_{T\in\mathcal{C}} \sum_{i \in \mathcal{L}} \log_2 Q_T(i)\label{eq:reformulation}
\end{equation}
Consider now an arbitrary complete tree $T\in\mathcal{C}$. Since the tree is complete, there exist (at least) two leaves that are siblings, say $j$ and $j+1$. Denote by $k$ the corresponding branching node. The contribution of these two leaves to the objective function on the right-hand side of \eqref{eq:reformulation} can be written as
\begin{align}
\log_2 &Q_T(j) + \log_2 Q_T(j+1)\nonumber\\
&= \log [Q_T(k) q_0] + \log [Q_T(k) q_1]  \nonumber\\
&= \log Q_T(k) + \log Q_T(k) + \log q_0 + \log q_1.
\end{align}
Now consider the tree $T'$ that results from removing the nodes $j$ and $j+1$. The new set of leaf nodes is $\mathcal{L}'=k\cup\mathcal{L}\setminus\{j,j+1\}$ and the new set of branching nodes is $\mathcal{B}'=\mathcal{B}\setminus k$. Also $Q_T$ defines a weighted leaf distribution on $\mathcal{L}'$. The same procedure can be applied repeatedly by defining $T=T'$, until $T'$ consists only of the root node. We use this idea to re-write the objective function of the right-hand side of \eqref{eq:reformulation} as follows.
\begin{align}
\sum_{i\in\mathcal{L}}&\log_2 Q_T(i)\label{eq:objective}\nonumber\\
&=\sum_{i\in\mathcal{L}'}\log_2 Q_T(i)+\log_2 Q_T(k) + \log_2 q_0+\log_2 q_1\nonumber\\
&=\sum_{k\in\mathcal{B}} \log_2 Q_T(k)+(2^m-1)[\log_2 q_0+\log_2 q_1].
\end{align}
Since $(2^m-1)[\log_2 q_0+\log_2 q_1]$ is a constant independent of the tree $T$, we have
\begin{align}
\argmax_{T \in \mathcal{C}} \sum_{i \in \mathcal{L}} \log_2 Q_T(i)=\argmax_{T \in \mathcal{C}}\sum_{k\in\mathcal{B}} \log_2 Q_T(k).\label{eq:reformulation2}
\end{align}
The right-hand side of \eqref{eq:reformulation2} is clearly maximized by the complete tree with the branching nodes with the greatest weighted probabilities. According to \cite[p. 47]{masseyapplied1}, this is exactly the tree that is constructed when Tunstall coding is applied to the weighted distribution $Q$.

\emph{Part ii.} We now consider $q_0\leq 1$ and $q_1\leq 1$. Assume we have constructed a non-complete binary tree. Because of non-completeness, we can remove a branch from the tree. Without loss of generality, assume that this branch is labeled by a zero. Denote by $\mathcal{S}$ the leaves on the subtree of the branch. Denote the tree after removing the branch by $T'$. Now,
\begin{align}
Q_{T'}(i)=\frac{Q_T(i)}{q_0}\geq Q_T(i), \text{ for each }i\in\mathcal{S}
\end{align}
where the inequality follows because by assumption $q_0\leq 1$. Thus, for the new tree $T'$, the objective function \eqref{eq:objective} is bounded as
\begin{align}
\sum_{i\in\mathcal{L}}\log_2 Q_{T'}(i)&=\sum_{i\in\mathcal{L}\setminus S}\log_2 Q_T(i)+\sum_{i\in\mathcal{S}}\log_2 \frac{Q_T(i)}{q_0}\nonumber\\
&\geq\sum_{i\in\mathcal{L}}\log_2 Q_T(i).
\end{align}
In summary, under the assumption $q_0\leq 1$ and $q_1\leq 1$, the objective function \eqref{eq:objective} that we want to maximize does not decrease when removing branches, which shows that there is an optimal complete tree. This proves the statement ii. of the proposition.
\end{IEEEproof}
\section{Minimizing I-divergence Per Bit}\label{sec:ninfodiv}
The following two propositions relate the problem of minimizing the I-divergence per bit to the problem of minimizing the un-normalized I-divergence.

Let $\mathcal{T}'\subseteq\mathcal{T}$ be some set of binary trees with $2^m$ leaves and define
\begin{align}
\Delta:=\min_{T\in \mathcal{T}'}\frac{\kl(U_T\Vert Q_T)}{\expop_{U_T}(L)}.\label{eq:defmin}
\end{align}

\begin{proposition}\label{prop:equivalenceNormalized}
We have
\begin{equation}
T^*:=\argmin_{T\in\mathcal{T}'} \frac{\kl(U_T\Vert Q_T)}{\expop_{U_T}(L)} = \argmin_{T\in\mathcal{T}'} \kl(U_T\Vert Q^\Delta_T) \label{eq:minf2v}
\end{equation}
where $Q^\Delta_T$ is the weighted distribution induced by $Q\circ 2^\Delta:=[q_02^\Delta,q_12^\Delta]$. 
\end{proposition}
\begin{IEEEproof}
By \eqref{eq:defmin}, for any tree $T\in\mathcal{T}'$, we have
\begin{align}
&\frac{\kl(U_T\Vert Q_T)}{\expop_{U_T}(L)} \geq \Delta\text{ with equality if } T = T^* \\
\Rightarrow&\kl(U_T\Vert Q_T) - \Delta\expop_{U_T}(L) \geq 0 \nonumber\\&\hspace{2cm}\text{ with equality if } T = T^*\label{eq:normequiv1}
\end{align}
We write the left-hand side of \eqref{eq:normequiv1} as
\begin{align}
\kl(U_T\Vert Q_T) &-\Delta\expop_{U_T}(L)\nonumber\\
&=\sum_{i\in\mathcal{L}}U_T(i)\log_2\frac{U_T(i)}{Q_T(i)}-\Delta\sum_{i\in\mathcal{L}}U_T(i)\ell_i\nonumber\\
&=\sum_{i\in\mathcal{L}}U_T(i)\left[\log_2\frac{U_T(i)}{Q_T(i)}-\log_2 2^{\Delta\ell_i}\right]\nonumber\\
&=\sum_{i\in\mathcal{L}}U_T(i)\log_2\frac{U_T(i)}{Q_T(i)2^{\Delta\ell_i}}.\label{eq:normequiv2}
\end{align}
Consider the path through the tree that ends at leaf $i$. Denote by $\ell_i^0$ and $\ell_i^1$ the number of times the labels $0$ and $1$ occur, respectively. The length of the path can be expressed as $\ell_i=\ell_i^0+\ell_i^1$. The term $Q_T(i)2^{\Delta\ell_i}$ can now be written as
\begin{align}
Q_T(i)2^{\Delta\ell_i}&=q_0^{\ell_i^0}q_1^{\ell_i^1}2^{\Delta(\ell_i^0+\ell_i^1)}\nonumber\\
&=(q_02^\Delta)^{\ell_i^0}(q_1 2^\Delta)^{\ell_i^1}\nonumber\\
&=Q^\Delta_T(i).\label{eq:deltainduced}
\end{align}
Using \eqref{eq:deltainduced} and \eqref{eq:normequiv2} in \eqref{eq:normequiv1} shows that for any binary tree $T\in\mathcal{T}'$ we have
\begin{align}
\sum_{i\in\mathcal{L}}U_T(i)\log_2\frac{U_T(i)}{Q^\Delta_T(i)}\geq 0\text{ with equality if } T=T^*
\end{align}
which is the statement of the proposition.
\end{IEEEproof}

\begin{proposition}\label{prop:norminfodiv}
Define
\begin{align}
\Delta:=\min_{T\in \mathcal{C}}\frac{\kl(U_T\Vert Q_T)}{\expop_{U_T}(L)}.
\end{align}
Then the optimal complete tree
\begin{align}
T^*:=\argmin_{T\in\mathcal{C}} \frac{\kl(U_T\Vert Q_T)}{\expop_{U_T}(L)}
\end{align}
is constructed by applying Tunstall coding to $Q_T^\Delta$.
\end{proposition}
\begin{IEEEproof}
The proposition is a consequence of Prop.~\ref{prop:equivalenceNormalized} and Prop.~\ref{prop:infodiv}.i.
\end{IEEEproof}
\subsection{Iterative Algorithm}
By Prop.~\ref{prop:norminfodiv}, if we know the I-divergence $\Delta$, then we can find $T^*$ by Tunstall coding. However, $\Delta$ is not known a priori. We solve this problem by iteratively applying Tunstall coding to $Q\circ 2^{\hat{\Delta}}$, where$\hat{\Delta}$ is an estimate of $\Delta$ and by updating our estimate. This procedure is stated in Alg.~\ref{alg:iterative}.
\begin{figure}
\begin{algorithm}\label{alg:iterative}\
\\
$\hat{T} \leftarrow \displaystyle\argmin_{T\in\mathcal{C}} \kl(U_T\Vert Q_T)$ \textcolor{blue}{\emph{solved by Tunstall coding on $Q$}}\\
\textbf{repeat} \\
\indent 1. $\hat{\Delta} = \frac{\kl(U_{\hat{T}}\Vert Q_{\hat{T}})}{\expop_{U_{\hat{T}}}(L)}$\\
\indent 2. $\hat{T} = \displaystyle\argmin_{T\in\mathcal{C}} \bigl[\kl(U_T\Vert Q_T)-\hat{\Delta}\expop_{U_T}(L)\bigr]$ \textcolor{blue}{\emph{Tunstall on $Q\circ 2^{\hat{\Delta}}$}}\\
\textbf{until} $\kl(U_{\hat{T}}\Vert Q_{\hat{T}})-\hat{\Delta}\expop_{U_{\hat{T}}}(L)=0$\\
$\Delta=\hat{\Delta}$, $T^*=\hat{T}$
\end{algorithm}
\vspace{-0.5cm}
\end{figure}
\begin{proposition}
Alg.~\ref{alg:iterative} finds $(\Delta,T^*)$ as defined in Prop.~\ref{prop:norminfodiv} in finitely many steps.
\end{proposition}
\begin{IEEEproof} The proof is similar to the proof of \cite[Prop.~4.1]{bocherer2012capacity}.

We first show that \emph{$\Delta$ is strictly monotonically decreasing}. Let $\hat{\Delta}_i$ be the value that is assigned to $\hat{\Delta}$ in step 1. of the $i$th iteration and denote by $\hat{T}_i$ the value that is assigned to $\hat{T}$ in step 2. of the $i$th iteration. Suppose that the algorithm does not terminate in the $i$th iteration. We have
\begin{align}
\hat{\Delta}_i&=\frac{\kl(U_{\hat{T}_{i-1}}\Vert Q_{\hat{T}_{i-1}})}{\expop_{U_{\hat{T}_{i-1}}}(L)}\nonumber\\
\Rightarrow &\kl(U_{\hat{T}_{i-1}}\Vert Q_{\hat{T}_{i-1}})-\hat{\Delta}_i\expop_{U_{\hat{T}_{i-1}}}(L)=0.
\end{align}
By step 2, we have
\begin{align}
\hat{T}_i = \argmin_{T\in\mathcal{C}} \bigl[\kl(U_T\Vert Q_T)-\hat{\Delta}_i\expop_{U_T}(L)\bigr]
\end{align}
and since by our assumption the algorithm does not terminate in the $i$th iteration, we have
\begin{align}
\kl(U_{\hat{T}_i}\Vert Q_{\hat{T}_i})-\hat{\Delta}_i\expop_{U_{\hat{T}_i}}(L)&<0\nonumber\\
\Rightarrow \frac{\kl(U_{\hat{T}_i}\Vert Q_{\hat{T}_i})}{\expop_{U_{\hat{T}_i}}(L)}&<\hat{\Delta}_i\nonumber\\
\Rightarrow \hat{\Delta}_{i+1}&<\hat{\Delta}_i.
\end{align}

Now assume the algorithm terminated, and let $\hat{T}$ be the tree after termination. Because of the assignments in steps 1. and 2., the terminating condition implies that for any tree $T\in\mathcal{C}$, we have
\begin{align}
\kl(U_T\Vert Q_T)-\hat{\Delta}\expop_{U_T}(L)\geq 0,\text{with equality if }T = \hat{T}.
\end{align}
Consequently, we have
\begin{align}
\frac{\kl(U_T\Vert Q_T)}{\expop_{U_T}(L)}\geq \hat{\Delta},\text{with equality if }T = \hat{T}.
\end{align}
We conclude that after termination, $(\Delta,\hat{T})$ is equal to the optimal tuple $(\Delta,T^*)$ in Prop.~\ref{prop:norminfodiv}. 

Finally, we have shown that $\Delta$ is strictly monotonically decreasing so that $\hat{T}_{i}\neq\hat{T}_{j}$ for all $i<j$. But there is only a finite number of complete binary trees with $2^m$ leaves. Thus, the algorithm terminates after finitely many steps.
\end{IEEEproof}
\section{Optimality of Complete Trees}\label{sec:complete}
\emph{Complete trees are not optimal in general: }
Consider $m=1$ and $Q\colon q_0=\frac{5}{6}, q_1=\frac{1}{6}$. For $m=1$, Tunstall coding constructs the (unique) complete binary tree $T$ with $2$ leaves, independent of which target vector we pass to it. The path lengths are $\ell_1=\ell_2=1$. The I-divergence per bit achieved by this is
\begin{align}
\frac{\kl(U_T\Vert Q_T)}{\expop_{U_T}(L)}&=\frac{-1-\frac{1}{2}\log_2(q_0 q_1)}{1}=0.424\text{ bits}.
\end{align}
Now, we could instead use a non-complete tree $T$ with the paths $0$ and $10$. In this case, I-divergence per bit is
\begin{align}
\frac{\kl(U_T\Vert Q_T)}{\expop_{U_T}(L)}&=\frac{-1-\frac{1}{2}\log_2(q_0 q_1q_0)}{\frac{1}{2}(1+2)}=0.37034\text{ bits}.
\end{align}
In summary, for the considered example, using a complete tree is sub-optimal. We will in the following derive simple conditions on the target vector $Q$ that guarantee that the optimal tree is complete.
\subsection{Sufficient Conditions for Optimality}
\begin{proposition}
Let $Q$ be a distribution. If $\max\{q_0,q_1\} \leq 4\min\{q_0,q_1\}$, then the optimal tree is complete for any $m\geq 1$ and it is constructed by Alg.~\ref{alg:iterative}.
\end{proposition}
\begin{IEEEproof}
According to Prop.~\ref{prop:infodiv}.ii, the tree that minimizes $\kl(U_T\Vert Q_T^\Delta)$ is complete if the entries of the weighted distribution $Q\circ 2^\Delta$ are both less than or equal to one. Without loss of generality, assume that $q_0\geq q_1$. Thus, we only need to check this condition for $q_0$. We have
\begin{align}
q_0 2^\Delta &\leq 1\nonumber\\
\Leftrightarrow \log_2 q_0 + \Delta &\leq 0\nonumber\\
\Leftrightarrow  \Delta &\leq -\log_2 q_0.\label{eq:condition}
\end{align}
We calculate the value of $\Delta$ that is achieved by the (unique) complete tree with $2$ leaves, namely
\begin{align}
\Delta=\frac{\kl(U_T\Vert Q_T)}{\expop_{U_T}(L)}=-1-\frac{1}{2}\log_2 (q_0 q_1).\label{eq:worstdelta}
\end{align}
For each $m\geq 1$, this $\Delta$ is achieved by the complete tree with all path lengths equal to $m$. Substituting the right-hand side of \eqref{eq:worstdelta} for $\Delta$ in \eqref{eq:condition}, we obtain
\begin{align}
-1-\frac{1}{2}\log_2 (q_0 q_1)&\leq-\log_2 q_0\nonumber\\
\Leftrightarrow1+\log_2(q_0 q_1)^{\frac{1}{2}}&\geq\log_2 q_0\nonumber\\
\Leftrightarrow2\sqrt{q_0 q_1}&\geq q_0\nonumber\\
\Leftrightarrow4 q_1 &\geq q_0
\end{align}
which is the condition stated in the proposition.
\end{IEEEproof}
\subsection{Asymptotic Achievability for Complete Trees}

\begin{proposition}\label{prop:perbitconvergence}
Denote by $T(m)$ the complete tree with $2^m$ leaves that is constructed by applying Alg.~\ref{alg:iterative} to a target distribution $Q$. Then we have
\begin{align}
\frac{\kl(U_{T(m)}\Vert Q_{T(m)})}{\expop_{U_{T(m)}}(L)}\leq\frac{\log_2\frac{1}{\min\{q_0,q_1\}}}{m}
\end{align}
and in particular, the I-divergence per bit approaches zero as $m\to\infty$.
\end{proposition}
\begin{IEEEproof}
The expected length can be bounded by the converse of the Coding Theorem for DMS \cite[p. 45]{masseyapplied1} as
\begin{align}
\expop_{U_{T(m)}}(L)&\geq\entop[U_T(m)]=m.
\end{align}
Thus, we have
\begin{align}
\frac{\kl(U_{T(m)}\Vert Q_{T(m)})}{\expop_{U_{T(m)}}(L)}\leq\frac{\displaystyle\min_{T'(m)\in\mathcal{C}}\kl(U_{T'(m)}\Vert Q_{T'(m)})}{m}.
\end{align}
The tree $T''(m)$ that minimizes the right-hand side is found by applying Tunstall coding to $Q$.
Without loss of generality, assume that $q_0\geq q_1$. According to the Tunstall Lemma \cite[p. 47]{masseyapplied1}, the induced leaf probability of a tree constructed by Tunstall coding is lower bounded as
\begin{align}
Q_{T''(m)}(i)\geq 2^{-m}q_1,\text{ for each leaf }i\in\mathcal{L}.
\end{align}
We can therefore bound the I-divergence as
\begin{align}
\kl(U_{T''(m)}\Vert Q_{T''(m)})&=\sum_{i\in\mathcal{L}}2^{-m}\log_2\frac{2^{-m}}{Q_{T''}(i)}\nonumber\\
&\leq\sum_{i\in\mathcal{L}}2^{-m}\log_2\frac{2^{-m}}{2^{-m}q_1}\nonumber\\
&=\log_2\frac{1}{q_1}.
\end{align}
We can now bound the I-divergence per bit as
\begin{align}
\frac{\kl(U_{T(m)}\Vert Q_{T(m)})}{\expop_{U_{T(m)}}(L)}&\leq\frac{\log_2\frac{1}{q_1}}{\expop_{U_{T(m)}}(L)}\leq\frac{\log_2\frac{1}{q_1}}{m}.
\end{align}
This proves the proposition. 
\end{IEEEproof}

\subsection{Optimality of Complete Trees for Large Enough $m$}

\begin{proposition}
For any target distribution $Q$ with $q_0 < 1$ and $q_1=1-q_0$, there is an $m_0$ such that for all $m>m_0$, the tree that minimizes 
\begin{align}
\frac{\kl(U_{T(m)}\Vert Q_{T(m)})}{\expop_{U_{T(m)}}(L)}
\end{align}
is complete.
\end{proposition}
\begin{IEEEproof}
Without loss of generality, assume that $q_0\geq q_1$. By Prop.~\ref{prop:perbitconvergence}, we have $\Delta_m\leq\frac{\log_2\frac{1}{q_1}}{m}$. Thus, there exists an $m_0$ such that
\begin{align}
q_12^{\Delta_m}\leq q_02^{\Delta_m}\leq q_02^{\frac{\log_2\frac{1}{q_1}}{m}}\leq 1, \text{ for all } m>m_0.
\end{align}
Thus, for all $m\geq m_0$, both entries of $Q_{T(m)}^{\Delta_m}$ are smaller than $1$. The proposition now follows by Prop.~\ref{prop:equivalenceNormalized} and Prop.~\ref{prop:infodiv}.ii.
\end{IEEEproof}
\begin{table}[ht!]
\centering
\caption{Comparison of v2f source coding and f2v distribution matching: $Q\colon q_0=0.615, q_1=0.385$; $m=2$}\label{tab:example}
\begin{tabular}{rcc}
&Tunstall on $Q$&Alg.~\ref{alg:iterative} on $Q$\\\addlinespace[0.1cm]\hline\addlinespace[0.1cm]
v2f source encoder&$\bsm{r} 00\mapsto 00\\01\mapsto 01\\10\mapsto 10\\11\mapsto 11\esm $&$\bsm{r} 1\mapsto 00\\01\mapsto 01\\001\mapsto 10\\000\mapsto 11\esm $\\
redundancy $\frac{\kl(\textcolor{red}{Q_T}\Vert U_T)}{m}$&0.038503&0.04176\\\addlinespace[0.1cm]\hline\addlinespace[0.1cm]
f2v distribution matcher&$\bsm{r} 00\mapsto 00\\01\mapsto 01\\10\mapsto 10\\11\mapsto 11\esm $&$\bsm{l} 00\mapsto 1\\01\mapsto 01\\10\mapsto 001\\11\mapsto 000\esm $\\
I-divergence per bit $\frac{\kl(U_T\Vert \textcolor{red}{Q_T})}{\expop_{U_T}(L)}$& 0.039206&0.037695
\end{tabular}
\vspace{-0.5cm}
\end{table}
\section{Source Coding Versus Distribution Matching}\label{sec:example}
An \emph{ideal} source encoder transforms the output of a DMS $Q$ into a sequence of bits that are independent and uniformly distributed. Reversely, applying the corresponding decoder to a sequence of uniformly distributed bits generates a sequence of symbols that are iid according to $Q$. This suggests to design a f2v distribution matcher by first calculating the optimal v2f source encoder. The inverse mapping is f2v and can be used as a distribution matcher. 

We illustrate by an example that this approach is sub-optimal in general. Consider the DMS $Q$ with
$Q\colon q_0=0.615, q_1=0.385.$
We calculate the optimal binary v2f source encoder with blocklength $m=2$ by applying Tunstall coding to $Q$. The resulting encoder is displayed in the 1st column of Table~\ref{tab:example}. Using the source decoder as a distribution matcher results in an I-divergence per bit of $0.039206$ bits. Next, we use Alg.~\ref{alg:iterative} to calculate the optimal f2v matcher for $Q$. The resulting mapping is displayed in the 2nd column of Table~\ref{tab:example}. The achieved I-divergence per bit is $0.037695$ bits, which is smaller than the value obtained by using the source decoder.

\emph{In general, the decoder of an optimal v2f source encoder is a sub-optimal f2v distribution matcher and the dematcher of an optimal v2f distribution matcher is a sub-optimal v2f source encoder.}

\bibliographystyle{IEEEtran}
\normalsize
\bibliography{IEEEabrv,confs-jrnls,references}

\end{document}